%% file: simex.tex
\documentclass[a4paper]{jpconf}
\pdfoutput=1
\usepackage{graphicx}
\usepackage[centertags]{amsmath}
\usepackage{amssymb}
\usepackage{latexsym}
\usepackage[]{glossaries}
\usepackage[caption=false]{subfig}
\usepackage[hidelinks]{hyperref}

\newlength{\picwidth}
\newlength{\eT}

\input{./glossary}

\makeglossaries
\begin{document}
\title{SIMEX: Simulation of Experiments at Advanced Light Sources}
\author{C Fortmann--Grote$^{1}$, A A Andreev$^{2,3}$, R Briggs$^{4}$, M Bussmann$^{5}$, A Buzmakov$^{6}$, M
Garten$^{5,7}$, A Grund$^{5,7}$, A Huebl$^{5,7}$, S Hauff$^{1}$, A Joy$^{8}$, Z
Jurek$^{9,10}$, N D Loh$^{11}$, T R\"uter$^{1}$, L
Samoylova$^{1}$, R Santra$^{9,10,12}$, E A Schneidmiller$^{13}$, A
Sharma$^{3}$, M Wing$^{8,13}$, S Yakubov$^{13}$, C H
Yoon$^{14}$, M V Yurkov$^{13}$, B Ziaja$^{9,10,15}$, A P Mancuso$^{1}$ }
\address{$^{1}$European XFEL GmbH, Holzkoppel 4, 22869 Schenefeld, Germany}
\address{$^{2}$Max Born Institute, Berlin, Germany}
\address{$^{3}$ELI ALPS, Szeged, Hungary}
\address{$^{4}$European Synchrotron Radiation Facility ESRF, Grenoble, France }
\address{$^{5}$Helmholtz--Zentrum Dresden--Rossendorf, Germany}
\address{$^{6}$Institute of Crystallography, Russian Academy of Sciences, Moscow, Russia}
\address{$^{7}$Technische Universit\"at Dresden, Germany}
\address{$^{8}$University College London, UK}
\address{$^{9}$Center for Free Electron Laser Science, Deutsches Elektronen Synchrotron, Hamburg, Germany}
\address{$^{10}$The Hamburg Center for Ultrafast Imaging, Hamburg, Germany}
\address{$^{11}$Department of Physics, National University of Singapore, Singapore}
\address{$^{12}$Department of Physics, University of Hamburg, Germany}
\address{$^{13}$Deutsches Elektronen Synchrotron, Hamburg, Germany}
\address{$^{14}$Linac Coherent Light Source, SLAC National Accelerator
Laboratory, Menlo Park, USA}
\address{$^{15}$Institute of Nuclear Physics, Polish Academy of Sciences, Krakow, Poland}
\ead{$^{1}$carsten.grote@xfel.eu}

\begin{abstract}
  \noindent
    Realistic simulations of experiments at large scale photon facilities, such as optical laser laboratories,
  synchrotrons, and free electron lasers, are of vital importance for the successful preparation, execution, and
  analysis of these experiments investigating ever more complex physical
  systems, e.g.\ biomolecules, complex
  materials, and ultra--short lived states of highly excited matter. Traditional photon science modelling
  takes into account only isolated aspects of an experiment, such as the beam
  propagation, the photon--matter
  interaction, or the scattering process, making idealized assumptions about the
  remaining parts, e.g.\ the
  source spectrum, temporal structure and coherence properties of the photon beam, or the detector response.
  In SIMEX, we have implemented a platform for complete start--to--end
  simulations, following the radiation from the source, through the beam transport optics to the sample or
  target under investigation, its interaction with and scattering from the sample, and its registration in a
  photon detector, including a realistic model of the detector response to the radiation. Data analysis tools
  can be hooked up to the modelling pipeline easily. This allows researchers and facility operators to simulate
  their experiments and instruments in real life scenarios, identify promising and unattainable
  regions of the parameter space and ultimately make better use of valuable beamtime.

  \settoheight{\eT}{T}
  {\centering\includegraphics[height=1.0\eT]{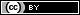}} This work is licensed under the Creative Commons Attribution 3.0 Unported
  License: http://creativecommons.org/licenses/by/3.0/.
\end{abstract}
\section{Introduction\label{sec:introduction}}
Simulations of photon science experiments that cover the complete path of the
photon from the source through beamline optics to the sample, its interaction
with and scattering from the sample, and its detection have recently been
used to model coherent diffraction
imaging experiments of single biomolecules at \AA ngstrom resolution \cite{Yoon2016}.
\Gls{spi} experiments (schematically depicted in Fig.~\ref{fig:spi_schema})
are performed at \glspl{xfel} that meet the stringent
requirements for ultra--short and intense x--ray pulses.
Such start--to--end simulations allow us to investigate the feasibility and
potential outcome of complex experiments under real--world conditions.
Every part of the experiment is simulated such that imperfections in the sources
spectral, temporal, and spatial structure, in the x--ray optical
components, radiation damage to the sample, and detector limitations are all
taken into account, allowing us to study how they affect the experimental observable and how these
error sources correlate with each other.

The computer program \texttt{simex\_platform} \cite{simex_github} was derived
from the \gls{spi} simulation suite \texttt{simS2E} \cite{Yoon2016}. Besides
\gls{spi}, \texttt{simex\_platform} supports simulation of various types of
experiments, involving various light sources (synchrotron sources, \glspl{fel},
optical lasers), various beamline optics, samples, and diagnostic
techniques. A python library provides
standardized user interfaces to the simulation codes and their parameters. This
approach was adopted from the \gls{ase}~\cite{Bahn2002, ase_www} which provides
python objects (so called \textit{Calculators}) that act as interfaces to ab--initio
electronic structure codes. Users can embed their own codes
into our simulation environment and run them under more realistic conditions compared to running them isolated
with idealized parameters and initial conditions.

Some simulation codes (called \textit{Backengines}) are readily shipped with \texttt{simex\_platform}, in
particular for \gls{spi} simulations. For other codes, we have
developed standardized data formats (data interfaces)
to integrate them into simulation workflows.
Our software has
modern and flexible deployment options including cmake, binary packages, and docker
containers packed for various operating systems \cite{Yakubov2016} and runs
in parallel on high--performance computing clusters.

\section{Concepts and structure of \texttt{simex\_platform}\label{sec:structure}}
\subsection{Simulation baseline\label{sec:baseline}}
Currently, \texttt{simex\_platform} supports simulation of
photon experiments that follow a sequential pattern involving five steps:
\begin{enumerate}
  \item Photon source: generation of radiation e.g.\ by accelerated charged
    particles or optical resonators.
  \item Photons propagate through a beamline consisting of apertures, slits, mirrors,
    gratings, and lenses.
  \item The photons interact with the sample.
  \item Scattered photons travel towards the detector.
  \item A detector registers incoming photons and produces a digital signal.
\end{enumerate}

\begin{figure}[t]
  \begin{center}
      \subfloat[Schematic of a \gls{xfel} single--particle imaging experiment. Photons emitted from the undulator (U) of the \gls{xfel} are deflected by offset
  mirrors (HOM) and focussed by focussing mirrors (FM) onto the sample (S).  Diffracted radiation is captured in the detector (D).]{%
      \includegraphics[width=0.49\textwidth,angle=0]{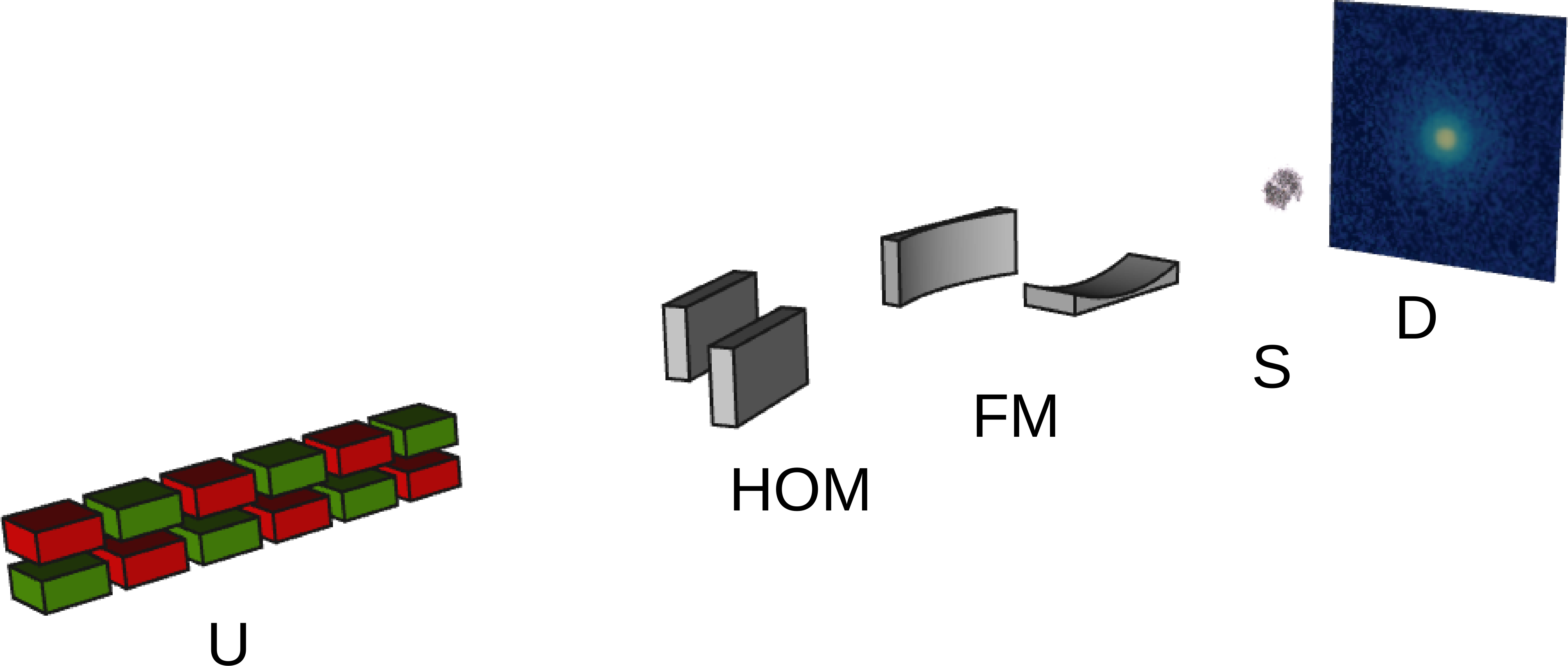}
      \label{fig:spi_schema}
        }
        \hspace{\fill}%
      \subfloat[Baseline workflow in
        \texttt{simex\_platform}. Blue boxes represent \textit{Calculators},
      green boxes are data interfaces.]{%
        \includegraphics[width=.45\textwidth,angle=0,clip, viewport= 0 110 790 500]{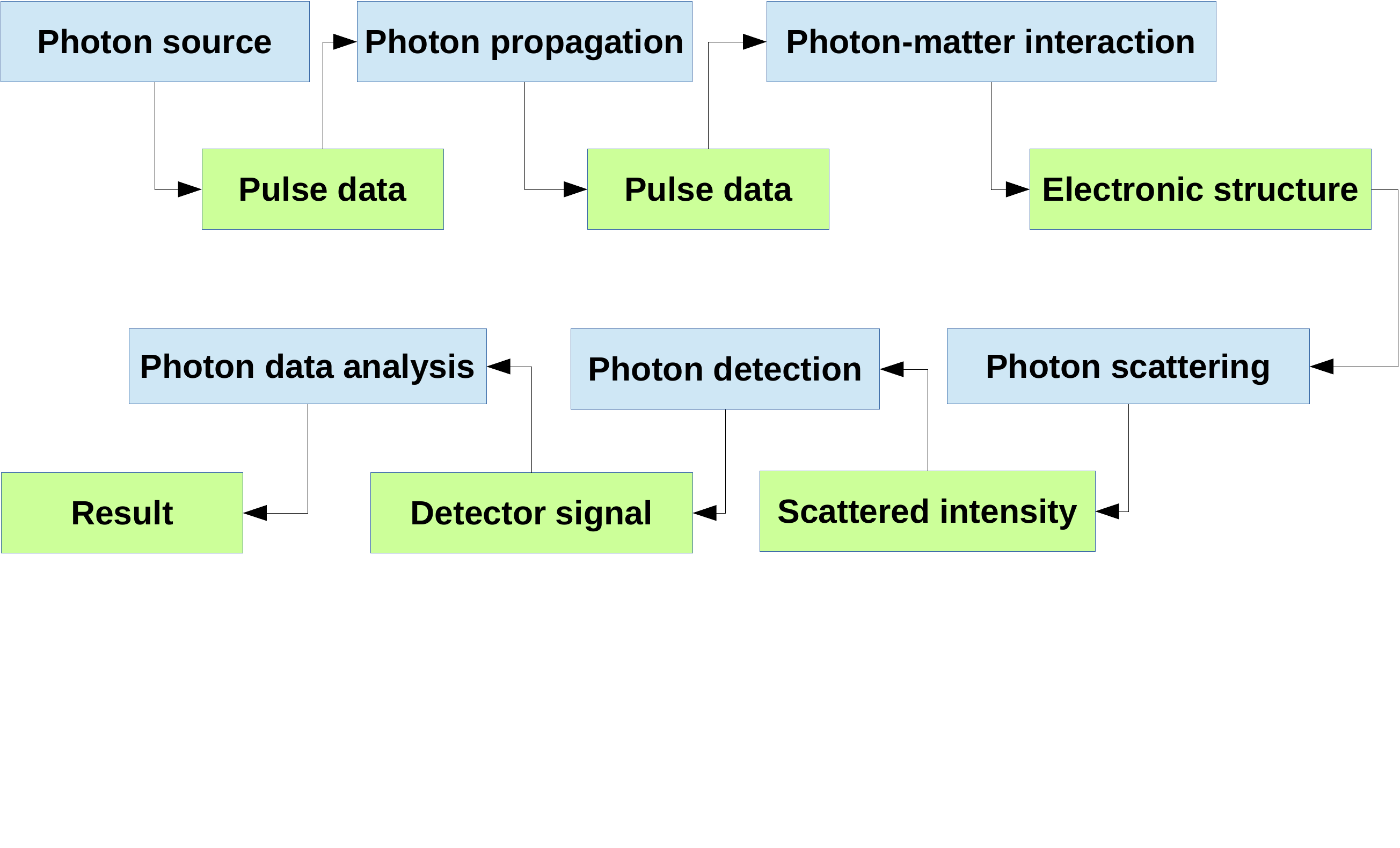}
        \label{fig:workflow}
      }
      \caption{}
  \end{center}
\end{figure}

\Gls{spi}, for which a schematic setup is shown in Fig.~\ref{fig:spi_schema} is
one example for a baseline application. The data flow between
\textit{Calculators} and intermediate data containers for which we have defined
standardized data hierarchies and formats is shown in Fig.~\ref{fig:workflow}.
\subsection{Calculators\label{sec:calculators}}
\begin{figure}[t]
  \begin{center}
    \centering{
      \includegraphics[width=.45\textwidth,angle=0,clip]{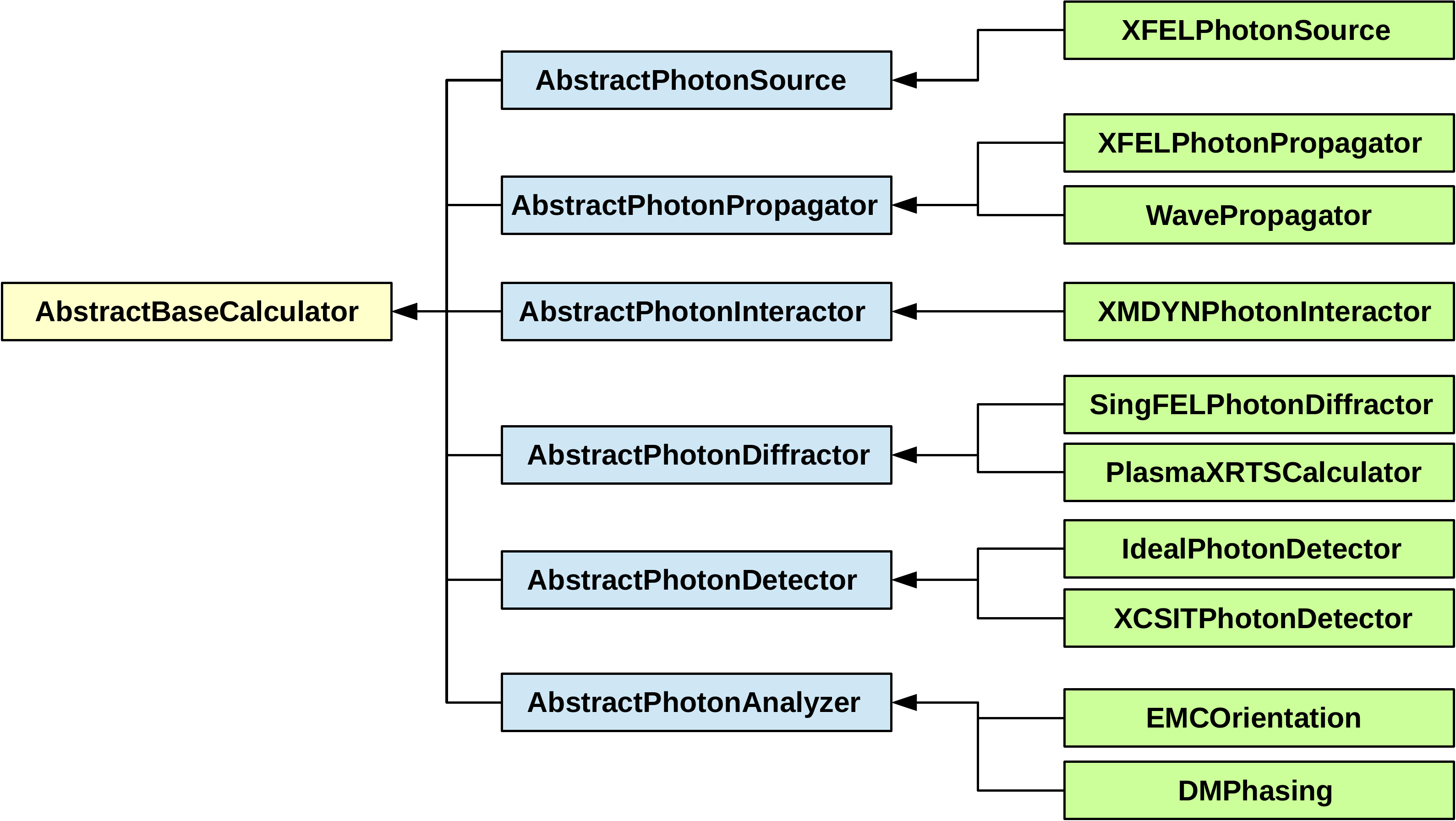}
    }
    \end{center}
    \caption{SIMEX \textit{Calculators} and their relationships}
  \label{fig:simex_oo_tree}
\end{figure}

Each of the five simulation tasks in a baseline application is represented by a suitable \textit{Calculator}.
\textit{Calculators} are organized in a
lightweight abstraction scheme consisting of three abstraction layers as depicted in
Fig.~\ref{fig:simex_oo_tree}

The top level class \textit{AbstractBaseCalculator} is a pure virtual object. It
declares virtual members that a deriving \textit{Calculator} must implement;
in this way we ensure basic functionality of the \textit{Calculator}, in
particular that \textit{Calculators} can exchange data and status information with each other.
These virtual members are a data container, where the data manipulated
by the \textit{Calculator} is stored, together with
corresponding query and assignment methods, furthermore methods for \textit{Backengine}
execution as well as \gls{io} utilities.
The intermediate level (blue boxes)
provides abstract base classes for each block of the simulation workflow:
AbstractPhotonSource, AbstractPhotonPropagator, AbstractPhotonInteractor,
AbstractPhotonDiffractor, and AbstractPhotonDetector. A sixth abstract
\textit{Calculator}, AbstractPhotonAnalyzer is provided to ease integration of
post--processing data analysis codes.

These classes are mainly responsible for declaring a list of variable names
expected from the previous \textit{Calculator} in the workflow and a list of variable
names provided for the next \textit{Calculator} in the workflow. Specialized \textit{Calculators}
adjust these lists according to their needs.

The third and final abstraction layer (green boxes) implements the
interfaces to the simulation codes
(\textit{Backengines}) that perform the actual numerical work on the input data.
They convert parameters and
input data into suitable formats for the \textit{Backengine},
before issuing the system or library calls,
gather the results after the calculation has finished, and write them
to an interface data file.

\subsection{Interfaces\label{sec:interfaces}}
Start--to--end simulations in \texttt{simex\_platform} require that any two
subsequent \textit{Calculators} employed in the simulation pipeline can
communicate data amongst each other. The data source has to write the data in a
format that the ensuing data sink can handle and interpret correctly. In
\texttt{simex\_platform}, we chose the Hierarchical Data Format (hdf5) \cite{HDFGroup1997}
as the underlying format for all simulation data files.  Data consistency in the
workflow is realized through the aforementioned mechanisms of exchanging
information about the required and provided data sets among the
\textit{Calculators}.
Recently, we have started
to adopt a more general approach of defining inter--calculator interfaces based
on the open standard for particle--mesh data \texttt{openPMD} \cite{openPMD}.
In particular, the transparent handling of physical units in \texttt{openPMD} makes it
possible to implement generic unit conversion mechanisms.

\section{Applications\label{sec:applications}}
As a first development milestone \texttt{simex\_platform} supports simulation of three
different baseline applications: single particle imaging at the European
\gls{xfel}, coherent x--ray scattering from high power laser excited plasmas and
x--ray probing of warm dense matter produced by high energy laser shock
compression.
\subsection{Single particle imaging\label{sec:spi}}
Femtosecond coherent x--ray pulses are generated in the SASE1 beamline of
the European \gls{xfel}. The \gls{sase} process is simulated with the code
\texttt{FAST}
\cite{Saldin1999}, which gives the electric field distribution at the exit of
the undulator section in a plane perpendicular to the beam direction as a
function of time (discrete time slices). \texttt{FAST} is not public,
we use precomputed datasets publicly available from a web database
\cite{xpd_xfel}, instead.

The pulses are then propagated through the beamline and the
SPB--SFX instrument's  \cite{Mancuso2013} focussing optics \cite{Bean2016} to the sample interaction point.
Propagation is modelled with the python library \texttt{WPG} \cite{Samoylova2016,
wpg_github} with realistic parameters for the various optical components in the
beamline. The output is again the electric field distribution as a function of
time in the sample interaction plane transverse to the beam direction.

The ultra--short and intense x--ray pulses interact with the sample primarily via
photo--ionization processes. Secondary Auger electrons quickly launch an
ionization cascade creating a nano--plasma that starts to expand on timescales of
10--100 fs after the first photons hit the sample. These processes are calculated
by means of \gls{md} simulations with the code \texttt{XMDYN} \cite{Jurek2016},
coupled to a \gls{mc} engine that
describes the photon--matter interaction processes in a stochastic manner. Rates
and cross sections are read from tabulated Slater--Hartree--Fock calculations
using the code \texttt{XATOM} \cite{Jurek2016, Son2011}. We save the entire sample trajectory (atom
positions and scattering form factors). We note here in passing that earlier
simulations of \gls{spi} described in Ref.~\cite{Ayyer2015, Serkez2014}
completely neglect radiation damage.

Scattering from the biomolecule is simulated  in the far
field approximation with the code \texttt{SingFEL} \cite{singfel_repo}.
\texttt{SingFEL} utilizes the output from the photon--matter interaction code to
calculate the instantaneously scattered light intensity by multiplying the
intensity at the sample with the appropriately weighted form factors and
geometry factors such as the detector solid angle. The instantaneous diffraction
is integrated over the pulse duration and saved as a 2D array representing the
pixel area detector.

The detector response to the incoming diffracted radiation can be calculated
with the \gls{xcsit} software \cite{Joy2015}. It describes particle generation,
charge transport, and electronic signal processing.
This can be used to simulated various detector effects like
variations in pixel performance, non--linear gain, electronic noise, and counting statistics.

The detector simulation concludes the simulation pipeline. In \gls{spi}, diffraction patterns are submitted to a sequence of
data post--processing algorithms to reconstruct the electron density from the measured diffraction data. \texttt{simex\_platform} provides
interfaces and \textit{Calculators} for orientation and phasing algorithms described in \cite{Loh2009}.
\subsection{Scattering from hot dense plasmas\label{sec:hpl}}
Hot (few hundreds of eV) and dense (near solid density and beyond) plasmas are
generated during the interaction of ultra--intense ($>1\times
10^{17}\,\text{W}/\,\text{cm}^{2}$) ultra--short (of the order $10\,\text{fs}$)
optical laser pulses with solid targets (e.g.\ metal foils). The plasma is
characterized by strong density and temperature gradients and
plasma instabilities. These features can be characterized by coherent x--ray scattering
\cite{Kluge2016}.
The photon source and propagation are described in the same way as
for the \gls{spi} example above (Sec.~\ref{sec:spi}).
The short--pulse
optical laser--plasma interaction is modelled with \texttt{PIConGPU}
\cite{Bussmann2013, picongpu_github}, an open source, explicit,
relativistic 3D \gls{pic} code employing
finite difference time domain techniques~\cite{Yee1966} to solve the Maxwell
equations coupling the laser field and the plasma particles.
For x--ray scattering calculations, we describe the radiation by a
photon distribution, which we convert from the field distribution delivered by
the propagation \textit{Calculator}. The simulation tracks the photons through the plasma volume
simulated by \texttt{PIConGPU} using the software \texttt{paraTAXIS}. The distribution of
scattered photons across the detector plane can then be fed into the detector
simulation as described in Sec.~\ref{sec:spi}.
\subsection{X--ray absorption and radiography in warm dense matter\label{sec:wdm}}
High energy (few tens of J) pulses of optical laser light impinge on a solid
target surface creating a shock wave travelling through the target. Pulse
shaping and multiple shock compression can create density, pressure, and
temperature conditions similar to planetary interiors \cite{Ping2006}, also referred to
as \gls{wdm}.  The target is probed by x--ray absorption spectroscopy and
radiography to measure the thermodynamic conditions before, during, and after
traversal of the shock wave.

In this case, the laser--matter
interaction is modelled with radiation--hydrodynamic codes, solving the
partial differential equations of radiation transport and hydrodynamic plasma
motion. Two codes are currently interfaced,
the 1D  code ``Esther''
\cite{Colombier2005} and the 2D code ``MULTI2D'' \cite{Ramis2009}. The most important
variables considered in these codes are those associated with the extreme
conditions generated by the high--power lasers: pressure (density), temperature,
and velocity. Feedback from these hydrocode outputs are crucial in the design and
implementation of laser shock experiments with x--ray interactions.

By measuring the x--ray absorption as a
function of position (radiography) and photon energy (\gls{xafs}), the shock
front structure, shock propagation dynamics, and ionic structure of the target
can be monitored as a function of time by varying the delay between optical pump
pulse and the x--ray probe.

The \gls{xafs} signal is a measure of the crystalline structure or liquid near
order in the shock
compressed target making it possible to identify structural phase transitions induced by the
shock.  XAFS can be simulated by combining first principle electronic
structure methods with linear response theory. A well known implementation is
the non--open source code FEFF \cite{Rehr2010}.

Radiography can be used to image e.g.\ the shock front during dynamical
compression. The Oasys \cite{Rio2014} framework has the capability to calculate
radiographs from simulated density profiles.

\section{Summary and Outlook\label{sec:summary}}
We have presented in this work an open--source platform for simulation of
experiments at advanced laser light sources. Our python library provides user
interfaces and data format conventions that enable simulations of entire
experiments from source to detector. \textit{Calculator} interfaces to
various simulation codes, both open and closed source, already exist,
particularly for \gls{spi} simulations. For other applications, x--ray
scattering from short--pulse laser excited plasmas and warm dense matter
diagnostics, we have collected the simulation codes and user interfaces will be
provided in the next development steps.

Data format definitions are derived from the openPMD standard to facilitate
communication between subsequent \textit{Calculators}. These data interfaces will also
enable us to perform simulations that go beyond the baseline scheme. Examples
are simulation of coherent light sources based on laser--plasma accelerated
electron beams and the generation of ion beams through laser--plasma interaction,
their interaction with matter, and medical applications.

\section*{Acknowledgements}
CFG, RB, AH, and SY acknowledge support from the European Cluster of Advanced Laser Light Sources
(EUCALL) project which has received funding from the \textit{European Union’s Horizon 2020
research and innovation programme} under grant agreement No 654220.
MW acknowledges support from the Alexander--von--Humboldt Foundation.
%
% Bibligraphy
%
%\printbibliography[]
\section*{References}
\newcommand*{\doi}[1]{\href{http://dx.doi.org/\detokenize{#1}{\raggedright\detokenize{#1}}}}

\end{document}

%% file: glossary.tex
\newacronym{ol}{OL}{optical laser}
\newacronym{xfel}{XFEL}{x-ray free-electron laser}
\newacronym{fel}{FEL}{free-electron laser}
\newacronym{esrf}{ESRF}{European Synchrotron Radiation Facility}
\newacronym{desy}{DESY}{Deutsches Elektronen-Synchrotron}
\newacronym{slac}{SLAC}{Stanford Linear Accelerator Center}
\newacronym{lcls}{LCLS}{Linac Coherent Light Source}
\newacronym{md}{MD}{Molecular Dynamics}
\newacronym{mc}{MC}{Monte Carlo}
\newacronym{pic}{PIC}{Particle In Cell}
\newacronym{wpg}{WPG}{WavePropaGator}
\newacronym{srw}{SRW}{Synchrotron Radiation Workshop}
\newacronym{simex}{SIMEX}{Simulation of Experiments}
\newacronym{eucall}{EUCALL}{European Cluster of Advanced Laser Light Sources}
\newacronym{hpl}{HPL}{high power laser}
\newacronym{wdm}{WDM}{warm dense matter}
\newacronym{hed}{HED}{high energy density}
\newacronym{eli}{ELI}{Extreme Light Infrastructure}
\newacronym{nif}{NIF}{National Ignition Facility}
\newacronym{lmj}{LMJ}{Laser Mega-Joule}
\newacronym{spi}{SPI}{single particle imaging}
\newacronym{ase}{ASE}{Atomic Simulation Environment}
\newacronym{sr}{SR}{Synchrotron Radiation}
\newacronym{hpc}{HPC}{high performance computing}
\newacronym{io}{IO}{input-output}
\newacronym{xafs}{XAFS}{x-ray absorption finestructure spectroscopy}
\newacronym{xcsit}{\texttt{X-CSIT}}{x-ray camera simulation toolkit}
\newacronym{sase}{SASE}{self-amplification of spontaneous emission}